\documentclass[prb,twocolumn]{revtex4}

\usepackage{amsmath,amsfonts}
\usepackage{graphicx}
\usepackage{enumerate}

\newcommand{\e}				{\mathrm{e}}
\newcommand{\dx}			{\mathrm{d}x}
\newcommand{\dy}			{\mathrm{d}y}
\newcommand{\dt}			{\mathrm{d}t}
\newcommand{\sump}			{\sideset{}{'}{\sum}}
\newcommand{\mmean}[1]		{\left\langle #1\right\rangle}
\newcommand{\mean}[1]		{\langle #1\rangle}

\begin{document}

% -------------------------------------------------------------------

\title{Rectification in one--dimensional electronic systems}

\author{Bernd Braunecker, D. E. Feldman, and J. B. Marston}
\affiliation{Department of Physics, Brown University, Providence, RI 02912}

\date{\today}

\pacs{73.63.Nm,71.10.Pm,73.40.Ei}

\keywords{Luttinger liquid; rectification; quantum wires}

% -------------------------------------------------------------------

\begin{abstract}
Asymmetric current--voltage ($I(V)$) curves, known as the diode or rectification effect, 
%,known as the diode or ratchet effect,
in one--dimensional electronic conductors can have their origin from scattering
off a single asymmetric impurity in the system. 
We investigate this effect in the framework of the Tomonaga--Luttinger model
for electrons with spin. 
We show that electron interactions strongly enhance the diode effect and lead to a 
pronounced current rectification even if the impurity potential is weak.
For strongly interacting electrons and not too small voltages, 
the rectification current, $I_r = [I(V)+I(-V)]/2$, measuring the asymmetry
in the current--voltage curve, has a power--law
dependence on the voltage with a negative exponent, $I_r \sim V^{-|z|}$,
leading to a bump in the current--voltage curve.
\end{abstract}

% -------------------------------------------------------------------

\maketitle

% -----------------------------------------------------------------------------

\section{Introduction}
Current rectification, also known as the diode effect,
occurs whenever the transport characteristics in electronic 
conductors are asymmetric under the application of the driving
voltage.
This asymmetry can have various origins. Best known perhaps
is the mismatch of band structures in macroscopic diodes, 
which leads to a potential wall that blocks the motion of
the particles in one direction, while it is not seen by 
the particles moving in the opposite direction. 

In the last years, rectification in mesoscopic and 
single--molecule systems has attracted much 
attention. Even though rectification by 
single asymmetric molecules has been suggested 30 years 
ago\cite{Aviram74}, it has been realized experimentally
in the 1990s only\cite{Geddes90, Martin93,Joachim00}. 
Asymmetric wave guides were constructed in the inversion
layer of semiconductor heterostructures\cite{Linke99,Lofgren03}.
Transport asymmetries have been observed in carbon 
nanotubes\cite{Postma01,Papadopoulos04}
and in tunneling in quantum Hall edge states\cite{Roddaro03}.
The experimental progress has been accompanied by much theoretical 
activity\cite{Christen96,Reimann97,Lehmann02,Scheidl02,Komnik03,Sanchez04,Spivak04,Feldman05}
with the main focus on Fermi--liquid systems.

\begin{figure}
\begin{center}
	\includegraphics[width=0.8\columnwidth]{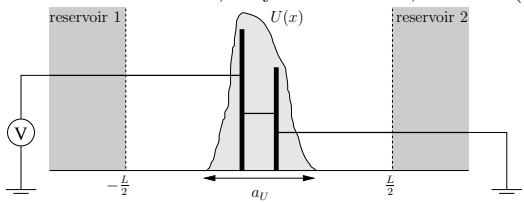}
	\caption{
	Sketch of system of length $L$ with a simple asymmetric potential $U(x)$,
	consisting in two point scatterers of different magnitude in a region 
	$a_U \ll L$. 
	The horizontal lines qualitatively represent the density profile 
	(averaged over Friedel oscillations) expected in the system.
	The reservoirs determine the densities of the injected particles 
	at the far left and 
	right sides, and are controlled by the voltage source.	
	The shaded area represents a more general asymmetric impurity
	potential, which would have a similar asymmetric density profile.
	\label{fig:system}}
\end{center}
\end{figure}

\begin{figure}
\begin{center}
	\includegraphics[width=0.8\columnwidth]{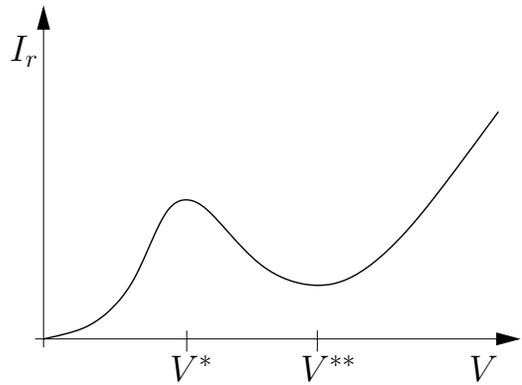}
	\caption{
	Qualitative behavior of the rectification current, $I_r(V) = [I(V)+I(-V)]/2$, 
	in the region,
	where the leading (third order) exponent is negative, 
	$z<0$. For voltages larger than
	$V^{*}$, the rectification current decreases until, at $V^{**}$,
	the leading second order term exceeds the third order amplitude,
	and leads to a further increase of the current.
	\label{fig:upturn}}
\end{center}
\end{figure}

An interesting source for rectification arises when the
particle interaction is strong enough that a single--particle
description becomes invalid. This naturally occurs in one--dimensional
systems, for which a Luttinger liquid behavior is expected.
It has been known for some time now\cite{Kane92} that in such systems the current
is strongly affected by the presence of impurities, even when they 
are weak, due to their renormalization by the electron--electron
interactions. 
One can thus expect that asymmetries in the impurity distribution 
leads to strongly asymmetric current--voltage curves. 
This issue was addressed very recently in the framework
of Luttinger liquids of spinless particles\cite{Feldman05}.
The \emph{rectification current} $I_r(V)=[I(V)+I(-V)]/2$ 
can be measured as the dc response to a 
low-frequency square voltage wave of amplitude $V$ and
expresses the asymmetry of the 
$I-V$ curve for forward and reverse bias. 
It was shown that a single weak asymmetric impurity is sufficient 
for a pronounced rectification effect (see Fig.~\ref{fig:system} for
a sketch of the system), leading to a large rectification current,
$I_r(V)$, at low voltages $V$.
Moreover an unusual behavior of the current was revealed for 
systems with strong repulsive interactions: 
A power--law dependence of $I_r$ on the voltage with a negative
exponent, $I_r \sim V^{-|z|}$, within a range $V^* < V < V^{**}$
[with $V^*$ and $V^{**}$ being expressed by some powers of the impurity
potential $U$; see also below], i.e. the rectification current \emph{increases} 
as the voltage is lowered. At $V<V^*$ the increase 
crosses over into a regular decrease such that the equilibrium 
condition $I=0$ at $V=0$ is met. The qualitative behavior is shown in 
Fig.~\ref{fig:upturn}.

\begin{figure}
\begin{center}
	\includegraphics[width=0.8\columnwidth]{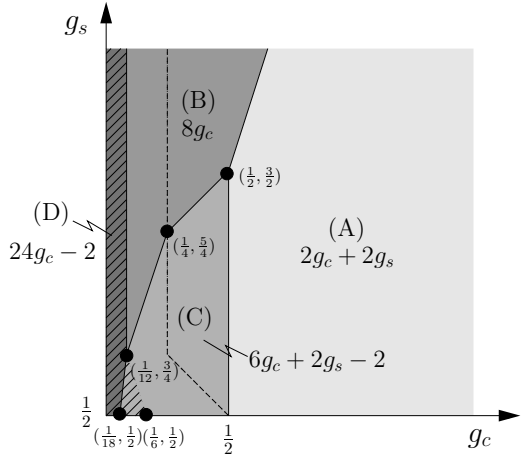}
	\caption{
	Diagram for the leading power--law contributions to the rectification
	current, $I_r \sim V^z$, for voltages $V \gtrsim V^*$. 
	The exponents $z$ for the four different
	phases are given by the formulae in the figure (see also Table~\ref{tbl:z}).
	The zone boundaries indicate crossover regions. 
	On the left of the dashed line $V^*>0$, and on the right $V^*=0$
	(see Eq.~\eqref{eq:V*}).
	In the hatched area, $z<0$, and the rectification current shows the 
	behavior sketched in Fig.~\ref{fig:upturn}.
	\label{fig:phases}}
\end{center}
\end{figure}

In this paper we pursue this work for real electrons carrying a spin.
We show that we can provide lower
and upper limits for the voltage $V$, within which perturbation theory 
can be applied and determine the leading power--law behavior of the rectification
current, $I_r \sim V^z$, as a function of the charge and spin 
interaction strengths, $g_c$ and $g_s$. This leads to a phase
diagram for the leading power--law dependence of $I_r$ with a 
much richer structure than in the spinless case (Fig.~\ref{fig:phases}).
We show that, similar to the spinless case, a negative exponent $z$,
$I_r \sim V^{-|z|}$, appears for strong repulsive electron--electron 
interactions within a voltage range that is determined by the
bare impurity scattering strength. Fig.~\ref{fig:phases} shows our
main result, the leading power--law behavior of the rectification current
$I_r$. 
The hatched region in the figure marks the range of 
$(g_c,g_s)$, in which the exponent $z$ becomes negative and in which the
rectification current shows the behavior sketched in Fig.~\ref{fig:upturn}.

The paper is organized as follows:
In the next section, we introduce the model of electrons in a 
one--dimensional system, and argue how scattering on a single impurity
can lead to rectification. We then quantitatively address this
problem within the bosonization approach, and show that 
the most relevant power--law behavior of the rectification current
can be obtained within second and third order perturbation theory. 
The results are listed in Table~\ref{tbl:z} and in Fig.~\ref{fig:phases}.
In the appendix we show that higher order perturbative contributions
cannot exceed the leading second and third order expressions.

% ----------------------------------------------------------------------------
\section{Model and Origin of Rectification}
\label{sec:model}

Let us consider a one--dimensional conductor of length $L$ with
electron--electron interactions that are effectively short--ranged
due to screening by gates. Such a system allows a description by the 
Tomonaga--Luttinger model, given by the Hamiltonian
\begin{align} 
	H 
	&= \int \dx \sum_\sigma \Bigl\{
	-\hbar v_F \Bigl[\psi_{R\sigma}^\dagger(x) i\partial_x \psi_{R\sigma}(x) 
	               - \psi_{L\sigma}^\dagger(x) i\partial_x \psi_{L\sigma}(x) \Bigr]
	\nonumber \\
	&\qquad \qquad \qquad \qquad
	+ U(x) \psi_{\sigma}^\dagger(x) \psi_{\sigma}(x)
	\Bigr\}
	\nonumber \\
	&+
	\int \dx \dy  \sum_{\sigma\sigma'}K_{\sigma\sigma'}(x-y) 
	\psi_{\sigma}^\dagger(x)\psi_{\sigma'}^\dagger(y)\psi_{\sigma'}(y)\psi_{\sigma}(x),
\label{eq:Hamiltonian}
\end{align}
where $\psi_{R\sigma}$ and $\psi_{L\sigma}$ are the operators for
right-- and left--moving electrons with spin $\sigma=\uparrow,\downarrow$, and
$\psi_\sigma = \psi_{R\sigma} + \psi_{L\sigma}$ is the conventional electron operator.
$U_\sigma(x)$ is the asymmetric potential (i.e. $U(x) \neq U(-x)$) 
localized in a small region about $x=0$.
$K(x-y)$ describes the electron--electron interaction. 
For the following discussion it is important to assume that the
long--ranged Coulomb interaction is screened by the gates, so that $K(x-y)$
becomes a short--ranged, rapidly decreasing function of $(x-y)$.

On its two ends, at $x=\pm L/2$, the system is adiabatically coupled to electrodes that 
serve as reservoirs of particles, and whose chemical potentials $\mu_{1,2}$ 
are controlled experimentally:
We assume that one electrode is fixed at the ground,
$\mu_2 = \mu$, while the other one is connected to the voltage source,
$\mu_1 = \mu + eV$. For such situations, it is possible to distinguish
between two effects, addressed more quantitatively below, that lead 
to rectification\cite{Feldman05}:

(1) \emph{The ``injected-density driven'' rectification} as 
the result of the dependence of the charge density on the voltage:
For simplicity, let us consider noninteracting particles.
For the voltages $\pm V$, only electrons in the energy ranges 
between $[\mu,\mu+eV]$ and $[\mu-eV,\mu]$ (at zero temperature) can
contribute to the transport. The presence of a scatterer $U$ 
with an energy--dependent transmission coefficient $R(E)$ in the system 
leads to different transmission coefficients for $\pm V$ and 
thus to different currents.
This is seen as follows:
For noninteracting particles, the reflection coefficient $R(E)$
is independent of the propagation direction\cite{LandauLifshitz},
and the current is
$I(V) \sim \int_{\mu_2}^{\mu_1} [1-R(E)] \mathrm{d}E$.
If the bandwidth $E_F$ is the only relevant scale for 
the energy dependence of $R(E)$, then, for small $U$ and $V$, we have 
$R(E) \sim U^2/E_F^2$ and the rectification current 
$I_r \sim \int_0^{eV} [R(E_F-E)-R(E_F+E)]\mathrm{d}E 
\approx 2 R'(E_F) \int_0^{eV} E \mathrm{d}E
\sim R(E_F) (eV)^2/E_F \sim U^2 (eV)^2/E_F^3$.
The rectification current is nonzero. For noninteracting particles it 
is proportional to $V^2$. As shown below, a modified power--law
dependence on $V$ is obtained in the presence of electron--electron
interactions.

We remark that this argument does not require a spatially asymmetric
impurity potential since the asymmetry is introduced through the 
injected charge 
densities.

(2) \emph{The ``asymmetric-impurity driven'' rectification effect} is independent of injected
densities.
It results from the renormalization of the asymmetric potential $U(x)$
by the electron--electron interactions, which leads to the asymmetric
current--voltage curves. For the Luttinger
liquid, this naturally involves multi--particle processes, so that all
such possible terms have to be taken into account in the modeling.
As shown in Ref.~\onlinecite{Feldman05}, this effect is absent
in the first two orders in the scattering potential $U$, and we
must address it perturbatively at the order $\sim U^3$.

The ``asymmetry-driven'' rectification effect can be qualitatively visualized 
(in a mean--field picture) as follows: In an interacting system,
electrons are backscattered by a combined potential $\tilde{U}(x) = U(x)+W(x)$,
where $W(x)$ is the self--consistent electrostatic potential created by 
the average local (nonuniform) charge density. 
Depending on the voltage sign, the density decreases or increases
as a function of the position $x$ and the magnitude of $U(x)$
(see Fig.~\ref{fig:system}).
Asymmetric impurities create different density profiles for 
opposite voltages, and lead to different combined backscattering potentials 
$\tilde{U}(x)$. The rectification effect is a consequence of the modification
of the current by the backscattering potential.

% ----------------------------------------------------------------------------

\section{Bosonization}
\label{sec:bosonization}

For the quantitative treatment of the rectification effect we use the bosonization technique,
which has become a standard tool for one--dimensional problems\cite{Giamarchi04}.
In the bosonization language, the creation of a right-- or left--moving electron
at the coordinate $x$ is expressed by
$\psi_{\nu\sigma}^\dagger(x) \sim \eta_{\nu\sigma}^\dagger \exp(+i [ \pm k_{F\nu} x + \phi_{\nu\sigma}(x)])$,
for $\nu = R,L$, where $k_{F\nu}/\pi$ are the densities of right-- and 
left--moving particles, where $\eta_{\nu\sigma}^\dagger$, the Klein factors,
raise the total number of right-- or left--moving electrons with spin $\sigma$ 
by one, and where $\phi_{\nu\sigma}(x)$ is a boson field that describes the dressing of 
this particle by a chain of particle--hole excitations. 
In correlation functions, the Klein factors keep track of particle conservation
and can contribute with signs to the expressions, as they anticommute
for different $(\nu,\sigma)$. These signs are all equal in the present calculations
since we assume conservation of the $S_z$ component of the spin by
the scattering process.
This allows us to drop the Klein factors in the following expressions.

In one dimension, the charge and spin degrees of freedom of the Hamiltonian
\eqref{eq:Hamiltonian} decouple, and it is convenient to set
$\phi_{c} = \sum_{\nu=L,R} [\phi_{\nu\uparrow}+\phi_{\nu\downarrow}]/2$
and 
$\phi_{s} = \sum_{\nu=L,R} [\phi_{\nu\uparrow}-\phi_{\nu\downarrow}]/2$,
which are bosonic fields related to the 
charge and spin densities as
$\rho_{c} = e [\partial_x \phi_{c} + 2 k_F] / \pi$ and
$\rho_{s} = (\hbar/2) \partial_x \phi_{s}/\pi$,
where $k_F = [k_{FR}+k_{FL}]/2$.

For $U=0$ the bosonization leads to a quadratic action 
for the fields $\phi_{c,s}$, which can be written in the form
\begin{equation} \label{eq:S_0}
	\mathcal{S}_0 = \int \dx \dt \sum_{a=c,s} \frac{1}{8\pi g_a}
	\left[(\partial_t \phi_a)^2 - (\partial_x \phi_a)^2 \right].
\end{equation}
The quantities $g_c$ and $g_s$ are the charge and spin interaction strengths
resulting from the screened electron--electron interaction $K(x-y)$. A noninteracting
system is characterized by $g_c = g_s = 1/2$. Repulsive (attractive) interactions
are expressed by $g_c < 1/2$ ($g_c > 1/2$). Interactions with conserved $SU(2)$ spin 
symmetry have $g_s = 1/2$.
For $g_s < 1/2$, a neglected sine--Gordon term of $\phi_s$ in $\mathcal{S}_0$,
describing spin--flip backscattering, would become relevant. 
The physics of such systems are considerably
different to the conductors described here (see e.g. Ref.~\onlinecite{Giamarchi04}), since
the fields $\phi_s$ would order and become massive. We exclude such situations explicitly by
assuming $g_s \ge 1/2$, such that all backscattering terms in $\mathcal{S}_0$ 
are scaled to zero.

If we assume that $U(x)$ is concentrated in a small region $a_U \ll L$ 
about $x=0$, the backscattering can be described by the values of the 
fields at $x=0$ only, $\Phi_{c,s}(t) \equiv \phi_{c,s}(x=0,t)$. 
This ``single--point'' description of $U(x)$, restricts the model to 
wavelengths longer than $a_U$,
i.e. to energy scales smaller than $E_U \ll \hbar v_F/a_U$ (with $v_F$ the 
Fermi velocity). Since the characteristic energy scale is set by the 
voltage, $E_U$ provides an upper limit for the applied voltage, $eV \ll E_U$,
and the validity of the model. 
Under this assumption, the impurity term has the form\cite{Kane92}, 
\begin{equation} \label{eq:S_imp}
	\mathcal{S}_{\text{imp}}
	=  
	- \int \dt \, 
	\sump_{n_c,n_s \in \mathbb{Z}} U(n_c,n_s) \e^{i \alpha(n_c,n_s)}
	\e^{i n_c \Phi_c + i n_s \Phi_s}.
\end{equation}
The integers $n_{c,s}$ physically express the transfer of $n_{c}$ charges
and $n_{s}$ spins between the right-- and the left--moving electrons
(where, for instance, $n_c=n_s=1$ corresponds to the backscattering
of an up--spin electron, and $n_c=0, n_s=2$ to the backscattering of an 
up--spin electron and a down--spin electron with opposite incident 
directions).
$U(n_c,n_s)$ and $\alpha(n_c,n_s)$ are modulus and phase of the effective
multi--particle backscattering potential (for symmetric potentials, $U(x)=U(-x)$,
the $\alpha(n_c,n_s)$ would vanish).
Forward scattering can be absorbed in $\mathcal{S}_0$ by a shift of $\phi_{c}$.
Since charges and spins are bound to physical electrons, the sum (indicated
by the $'$ next to it) runs over $n_{c,s}$ such that $n_c+n_s$ is even.
$|n_c|=|n_s|=1$ corresponds to the usual $2k_F$ backscattering of electrons.
The most relevant contribution to the currents arises from the 
backscattering processes with $n_c+n_s=0,\pm 2$. These terms dominate
in the ``density-driven'' rectification effect, but higher orders in $n_c+n_s$ 
are required for the ``asymmetry-driven'' rectification effect, and the
full sum over $n_{c,s}$, therefore, must be kept.

Throughout the following analysis, $U(x)$ is assumed to be weak, $|U| \ll E_F$ 
(with $E_F \sim$ bandwidth). Due to the strong renormalization
of the impurity strength through the electron--electron interactions,
the precise nature of $U$
is of no major importance; $U(n_c,n_s)$ is 
an effective many--particle potential, dressed by short--time
electron processes with frequencies well above $E_U$.
Its magnitude can be roughly estimated as\cite{Kane92}
\begin{equation} \label{eq:U}
	U(n_c,n_s) \sim k_F \int \dx \, \e^{2 i k_F n_c x} U(x),
\end{equation}
which means that $U(n_c,n_s) \sim U$.
Its further renormalization by low--energy processes, 
leading to a power--law correction in $V$,
is considered below. The coefficients $U(n_c,n_s)$ depend further
on the applied voltage through the densities $k_{FL}$ and
$k_{FR}$.

Since the Hamiltonian is Hermitian, $U(n_c,n_s) = U(-n_c,-n_s)$ and
$\alpha(n_c,n_s) = - \alpha(-n_c,-n_s)$. 
Because of the spin symmetry, $U(n_c,n_s) = U(n_c,-n_s)$ and 
$\alpha(n_c,n_s) = \alpha(n_c,-n_s)$.
For spatially asymmetric potentials, as considered here, 
$\alpha(n_c,n_s) \neq \alpha(-n_c,n_s)$.

If $U=0$, the system is characterized by right--moving particles
injected from the left reservoir with the chemical potential $\mu_1$,
and left--moving particles injected from the right reservoir 
with chemical potential $\mu_2$. Due to the absence of backscattering
in Eq.~\eqref{eq:S_0}, the
right-- and left--movers are in equilibrium with the corresponding
reservoirs.
In the presence of an impurity $U$, the equilibrium notion of 
chemical potentials becomes meaningless in the system; the quantities 
$\mu_{1,2}$ enter only through the voltage dependence of the injected 
carrier densities in the description and appear in the backscattering term
that couples right-- and left--moving particles. Explicitly, this can be seen
by switching to the interaction representation
$H \to \hat{H} = H - \mu_1 N_R  - \mu_2 N_L$ for $U=0$. 
This shifts the single particle energies of the injected particles,
and the electron operators become 
$\psi_R(x,t) \to \hat{\psi}_R(x,t) = \e^{i \mu_1 t/\hbar} \psi_R(x,t)$ and 
$\psi_L(x,t) \to \hat{\psi}_L(x,t) = \e^{i \mu_2 t/\hbar} \psi_L(x,t)$.
Hence, for $U\neq 0$, the backscattering of every charge described 
by $\mathcal{S}_{\text{imp}}$ acquires a time--dependence on the difference 
of chemical potentials, $\mu_1-\mu_2 = e V$, as $\e^{\pm i e V t/\hbar}$. 
For the $n_c$ backscattered charges, this becomes
\begin{equation} \label{eq:S_imp_int}
\begin{aligned}
	\mathcal{S}_{\text{imp}}
	=  
	- \int \dt \, 
	&\sump_{n_c,n_s \in \mathbb{Z}} 
	U(n_c,n_s) \e^{i \alpha(n_c,n_s)}
	\\
	&\times 
	\e^{i n_c \Phi_c + i n_s \Phi_s}
	\e^{i n_c e V t / \hbar}.
\end{aligned}
\end{equation}
The time--dependence of the backscattering operator reflects the 
nonequilibrium constraints. In order to calculate expectation values,
we have to consider an appropriate nonequilibrium technique, such 
as provided by the Keldysh method\cite{Keldysh65,RammerSmith86}:
We assume that in the remote past, $t\to -\infty$, the system is fully  
described by $\mathcal{S}_0$ and the ground state $|0\rangle$
of the shifted Hamiltonian $H-\mu_1 N_R -\mu_2 N_L$. 
Averages are taken over this well--defined ground state $|0\rangle$ 
only, while the impurity term $\mathcal{S}_{\text{imp}}$ is taken into
account through the time evolution operator $S(-\infty,t)$, given by
\begin{equation}
	S(t',t'') = T\exp\left(-\frac{i}{\hbar} \int_{t'}^{t''} \dt H_\text{imp}(t)\right),
\end{equation}
with $H_\text{imp}$ the impurity Hamiltonian in the interaction 
representation, and $T$ the time order operator. The expectation value
of an operator $O(t)$ is expressed by
\begin{equation}
	\mean{O(t)} = \langle 0 | S(-\infty,t) O(t) S(t,-\infty) |0\rangle,
\end{equation}
with $S(t,-\infty) = S(-\infty,t)^\dagger$. 

% ----------------------------------------------------------------------------
\section{Currents}
\label{sec:currents}

In a clean system, the current flow is proportional to the 
voltage, and is given by the Landauer formula\cite{Maslov95,Ponomarenko95,Safi95},
$I_0(V) = 2 V e^2/ h$ (the factor 2 accounts for the two
spin channels).
In the presence of the impurity scatterer $U$, the backscattering
corrects the current as $I(V) = I_0(V) + I_{bs}(V)$.
This shows that the rectification current depends only on the backscattering
currents,
$I_r(V) = [I_{bs}(V)+I_{bs}(-V)]/2$. 
The backscattering current operator can be obtained, for instance,
by the time variation of the number $N_R$ of right--moving particles,
$I_{bs} = \frac{\mathrm{d}}{\dt} N_R = i [H,N_R]/\hbar$.
If we set $e = \hbar = v_F = 1$ for the ease of notation, 
this yields, in the interaction representation, 
\begin{equation}
	I_{bs}(V,t) 
	= 
	i \sump_{n_c,n_s} n_{c} 
	U(n_c,n_s) \e^{i \alpha(n_c,n_s)}
	\e^{i n_c \Phi_c + i n_s \Phi_s} 
	\e^{i n_c V t},
\end{equation}
and we need to calculate 
\begin{equation} \label{eq:cur_av}
	\langle I_{bs}(V)\rangle
	= \langle 0| S(-\infty,0) I_{bs}(V,0) S(0,-\infty) | 0\rangle,
\end{equation}
where $S(t,t')$ is the time evolution operator, and $|0\rangle$ the ground state
of the system described by $\mathcal{S}_0$. 

% ----------------------------------------------------------------------------
\section{Results from weak coupling theory}
\label{sec:perturb}

For a weak impurity potential, the currents can be calculated within 
perturbation theory.

As shown by Kane and Fisher\cite{Kane92}, the 
backscattering potential is renormalized by the electron--electron
interactions and scales as a power--law with the characteristic 
energy (here set by $V$) as
$U_\text{eff}(n_c,n_s) \sim U(n_c,n_s) (V/E_F)^{g_c n_c^2+g_s n_s^2-1}$.
Perturbation theory is applicable when $U_\text{eff} \ll E_F$. 

For $n_c^2 g_c+n_s^2 g_s > 1$, this condition is always met for
$V,U \ll E_F$, while for $n_c^2g_c+n_s^2g_s < 1$, we must have
\begin{equation}  \label{eq:def_V*}
	V \gg V^*_{n_c,n_s} \equiv (U/E_F)^{1/(1-n_c^2g_c-n_s^2g_s)} E_F,
\end{equation}
for any admissible values of $n_c,n_s$. 
We set $V^*_{n_c,n_s}=0$ whenever $n_c^2 g_c + n_s^2 g_s > 1$.
Since $g_s \ge 1/2$, only $n_s = 0,1$ have to be considered. 
We always have  
$V^*_{2n+2,0} < V^*_{2 n,0}$ and 
$V^*_{2n+1,1} < V^*_{2n-1,1}$, so that the only important
quantities are $V^*_{1,1}$ and $V^*_{2,0}$. 
For $g_c +g_s < 1$ and $g_c < 1/4$ both $V^*_{1,1}$ and $V^*_{2,0}$
are nonzero, and we have
$V^*_{1,1} > V^*_{2,0}$ for $g_s < 3 g_c$.
The lower cutoff energy is given by 
\begin{equation} \label{eq:V*}
	V^* = \max[V^*_{1,1},V^*_{2,0}].
\end{equation}
Interestingly, if $V^* > 0$, perturbation theory requires a voltage
$V$ that is not too small, $V> V^*$, which is much in contrast to
the usual perturbation theory, in which $V$ would be required to 
be a small parameter close to $V=0$.

The upper energy limit is set by the energy $E_U$, at which the corrections
to the model become important. The perturbation theory is valid in 
the range $eV^* \ll eV \ll E_U$. 

In this case, the rectification current is dominated by the second and
third order perturbative expressions from an expansion in powers 
of $U$. Due to the renormalization of the potentials,
and due to the constraints on particle conservation and
even sums of $n_c+n_s$, third order expressions can become 
larger than the second order ones and have to be taken into
account. In the appendix we further show that, for $V>V^*$, 
higher orders
cannot exceed the second and third order expressions, and can
be safely neglected.

\subsection{Second order}

As mentioned, the rectification effect arises from
the asymmetry of the charge/spin density profiles. The
``injected-density-driven'' rectification effect is due to the dependence of the density
on the voltage, and hence depends on how the coefficients $U(n_c,n_s)$ 
change upon the variation of the density (see Eqs.~\eqref{eq:S_imp} and
\eqref{eq:U}).
The dominant contribution appears at second order in $U$, which 
reads after expanding Eq.~\eqref{eq:cur_av}
\begin{multline}
	\mean{I_{bs}^{(2)}(V)}
	=
	\int_{C_K} \dt 
	\sump_{n,m}
	i n_{c} \,
	U(n) U(m)
	\\
	\times
	\e^{i \alpha(n) + i \alpha(m)}
	\e^{i m_c V t}
	\mmean{ T_c \, 
	\e^{i n \Phi(0) }
	\
	\e^{i m \Phi(t) }
	}_{0},
\end{multline}
where we have used the notations $n = (n_c,n_s)$ and
$n\Phi(t) = n_c\Phi_c(t)+n_s\Phi_s(t)$.
$\mmean{...}_0$ is the average over the ground state of $\mathcal{S}_0$,
$C_K$ is the Keldysh contour $-\infty \to 0 \to -\infty$,
and $T_c$ is the chronological order operator on $C_K$.
Particle conservation imposes $m=-n$.
Since  $\alpha(-n) = -\alpha(n)$ the phases $\alpha$ cancel each other. 
Ignoring the 
voltage dependence of $U$, this expression, therefore, changes the sign
upon $V \to -V$, and would naively not contribute to the rectification 
effect (which is a consequence of the invariance of the action $\mathcal{S}_0$ 
under the change $\phi_{c,s} \to -\phi_{c,s}$). 
However, since $U$ depends on the voltage through the density $k_F$ 
(see Eq.~\eqref{eq:U}), we can expand it in powers of $V$ as
$U = U_0 + V U_1 + \dots$. 
At small voltages, $V \ll E_F$, the correction is linear, and
if the bandwidth $E_F$ is the only relevant scale for the energy 
dependence of $U$, we have
$U_1 \sim \tfrac{\mathrm{d}U_0}{\mathrm{d}E} \sim U_0 / E_F$.
The rectification current can be written as
(choosing $V>0$)
\begin{multline}
	I_r^{(2)}(V)
	= 2 V 
	\int_{C_K} \dt \sump_{n}
	i n_{c} \,
	U_0(n) U_1(n)
	\e^{-i n_c V t}
	\\
	\times
	\mmean{ T_c \, 
	\e^{i n \Phi(0) }
	\
	\e^{-i n \Phi(t) }
	}_{0}.
\end{multline}
The propagators in the latter equation have the usual form 
known from the Luttinger liquid theory\cite{Giamarchi04}:
\begin{multline}
	\mmean{ T_c \, 
	\e^{i n_c \Phi_c(0) + i n_s \Phi_s(0) }
	\
	\e^{-i n_c \Phi_c(t) - i n_s \Phi_s(t) }
	}_0
	\\
	=
	 \e^{n_c^2  \mmean{T_c \Phi_c(0) \Phi_c(t)}_0 
	    + n_s^2  \mmean{T_c \Phi_s(0) \Phi_s(t)}_0},
\end{multline}
where the free propagators are given by
\begin{equation}
	\mmean{\Phi_{c,s}(t_1) \Phi_{c,s}(t_2)}_0
	= - 2 g_{c,s} \ln\bigl(i(t_1-t_2) + \delta\bigr),
\end{equation}
with $\delta>0$ an infinitesimal constant.
We can now extract the voltage dependence of the current
by changing to the time variable $s = V t$.
The propagators become
\begin{multline}
	\mmean{ T_c \, 
	\e^{i n_c  \Phi_c(0) + i n_s \Phi_s(0) }
	\
	\e^{-i n_c \Phi_c(t) - i n_s \Phi_s(t) }
	}_0
	\\
	\sim
	V^{2 n_c^2 g_c + 2 n_s^2 g_s}.
\end{multline}
From the integration measure, we obtain another factor $V^{-1}$,
which cancels the $V$ from the expansion of the potential.
Hence the rectification current has the form
\begin{equation}
	I_r^{(2)}(V)
	= 
	\sump_{n_c,n_s} n_{c} C(n) V^{z^{(2)}(n)}
\end{equation}	
with $C$ constants of the order of $C \sim U_0 U_1 \cdot 1$,
and 
\begin{equation} \label{eq:z2}
	z^{(2)}(n) = z^{(2)}(n_c,n_s) = 2 n_c^2 g_c + 2 n_s^2 g_s.
\end{equation}
The leading order is obtained by minimizing the exponent, which
is achieved by one of the processes 
$(n_c,n_s) = (2,0), (0,2), (1,1)$ (and the combinations obtained
by changing signs) only, as backscattering of more particles leads 
to larger $z^{(2)}$.
Terms with $n_c=0$ do not couple to the voltage ($e^{i n_c V t} \equiv 1$) 
and coincide
with the equilibrium value at $V=0$. Therefore, their amplitude
vanishes. The comparison of the remaining two processes shows
the dominance of $(n_c,n_s)=(1,1)$ [labeled as (A) in Fig.~\ref{fig:phases}]
for $g_s < 3 g_c$ and $(n_c,n_s)=(2,0)$ for 
$g_s > 3 g_c$ [(B) in Fig.~\ref{fig:phases}].
The noninteracting system is characterized by $g_c = g_s = 1/2$, 
leading to $z=2$. This result certainly agrees with the result from the 
application of the Landauer formalism to a noninteracting problem.

\subsection{Third order}

The ``asymmetry-driven'' rectification effect appears at third order.
The contribution to the backscattering currents becomes
\begin{multline}
	\mean{I_{bs}^{(3)}(V)}
	=
	\frac{1}{2!}
	\int_{C_K} \dt_1 \dt_2 \sump_{n,m,l}
	i n_{c} \,
	\e^{i m_c V t_1 + i l_c V t_2}
	\\
	\times
	U(n) U(m) U(l)
	\e^{i \alpha(n) + i \alpha(m) + i \alpha(l)}
	\\
	\times
	\Bigl\langle T_c \, 
	\e^{i n \Phi(0)}
	\
	\e^{i m \Phi(t_1) }
	\
	\e^{i l \Phi(t_2) }
	\Bigr\rangle_{0},
\end{multline}
which has to be evaluated with the constraints 
$n_c + m_c + l_c = n_s + m_s + l_s = 0$, together with
$n_c + n_s, m_c + m_s$, and $l_c + l_s$ being even. 
Again, diagrams in which $n_c = m_c = l_c = 0$, do not couple to the 
voltage and vanish. 
The third order term is no longer invariant under the transformation
$(V,\Phi_c) \to (-V,-\Phi_c)$ because, generally, 
$\alpha(n)+\alpha(m)+\alpha(l) \neq 0$ (i.e.
$\alpha(n_c,n_s) + \alpha(m_c,m_s) \neq \alpha(n_c+m_c,n_s+m_s)$);
the invariance would require the additional transformation
$\alpha \to -\alpha$, corresponding to a mirror reflection $x\to -x$
of $U$.
Since the sum of the exponents $\alpha$ does not vanish, $I_r \neq 0$,
and the same analysis as before leads to
\begin{equation}
	I_r^{(3)}(V)
	=
	\sump_{n,m,l} 
	C(n,m,l)
	V^{z^{(3)}(n,m,l)},
\end{equation}
with 
\begin{equation} \label{eq:z3}
	z^{(3)}(n,m,l) =
	g_c (n_c^2 + m_c^2 + l_c^2) + g_s (n_s^2 + m_s^2 + l_s^2) - 2,
\end{equation}
and the amplitudes $C \sim U^3$. Table \ref{tbl:z} lists the 
smallest exponents $z$ at this order, characterized by the letters
(C) and (D), for which the rectification current is nonzero.
The exponent $z_D$ is smaller than $z_C$ for $g_s > 9 g_c$.

To obtain the leading power--law behavior for given $(g_c,g_s)$, 
we compare the amplitudes from second order, $\sim U^2 V^{z^{(2)}}$,
and third order, $\sim U^3 V^{z^{(3)}}$, for voltages
$V \gtrsim V^*$ using Eq.~\eqref{eq:def_V*}. If $V^*=0$, 
the mere comparison of exponents is sufficient.
The result of comparison is represented in Fig.~\ref{fig:phases}. 
The diagram is not a phase diagram in the 
proper sense because the crossover lines shift with the 
voltage. Close to the crossover lines, all contributions
of the neighboring phases are large, whereas far from the 
boundaries, a single power--law dominates.
Finally, as shown in the appendix, higher order perturbative 
corrections lead to less relevant power--laws and can be neglected. 

% ----------------------------------------------------------------------------
\section{Conclusions}

The above results show that the inclusion of spin degrees of 
freedom leads to a more diverse behavior of the rectification current
for different interaction strengths than in the spinless
case of Ref.~\onlinecite{Feldman05}. The leading power--laws
for the rectification current as a function of the interaction
strengths $g_c$ and $g_s$ are represented in Fig.~\ref{fig:phases}.

The most interesting region is characterized by a negative exponent,
$z<0$. It leads to the unusual behavior of a decreasing rectification 
current as the voltage $V > V^*$ is raised. Since negative exponents 
appear only from the third order contributions, they are
a realization of the ``asymmetry-driven'' rectification effect, and 
due to the presence of an asymmetric impurity in the system. 
The decrease stops at the upper voltage $V^{**} \sim U^{1/(|z^{(3)}|+z^{(2)})}$, 
where the amplitude of the second order contribution exceeds
that of the third order. The qualitative behavior is
sketched in Fig.~\ref{fig:upturn}.

The rectification effect is strong if the magnitude of $I_r$ becomes 
comparable to that of the total current $I(V) \sim V$ and to 
that of the most relevant contribution to the backscattering
current\cite{Kane92}, $I_{bs}(V) \sim V^{2 n_c^2 g_c+2 n_s^2 g_s-1}$
with $(n_c,n_s) = (1,1)$ or $(n_c,n_s)=(2,0)$.
For interaction strengths $g_{c,s}$ such that $V^*=0$,
the corrections are weak and generally $I_r(V) \ll I_{bs}(V) \ll I(V)$.
For $V^* > 0$, however, $I_{bs}(V)$ becomes
comparable to $I(V)$ at $V \gtrsim V^*$, i.e. close to the limit 
of validity of perturbation theory.
For such voltages,
in regions (C) and (D)
the ratio of rectification current to total current can 
roughly be estimated
as $I_r(V)/I(V)	\sim U^3 (V^*)^{z^{(3)}-1}$.
In phase (C), when $V^* = V^*_{1,1}$ (i.e. $3g_c - g_s > 0$), this leads
to the ratio 
$I_r(V)/I(V) \sim (V^*)^{3g_c-g_s} = U^{(3g_c-g_s)/(1-g_c-g_s)}$.
The rectification current, therefore, can become comparable to $I(V)$ 
(as well as to $I_{bs}(V)$)
when $3g_c-g_s$ is small. The comparison in the region where  
$V^*=V^*_{2,0}$ (i.e. $3 g_c - g_s < 0$)
yields a similar condition for $2g_s-6g_c$ being small.
In phase (D), we find $I_r(V)/I(V) \sim (V^*_{2,0})^{12g_c}$, 
which may lead to an enhancement of the rectification effect
for weak impurities as $g_c$ becomes small.

To conclude, we have shown that one--dimensional electronic
conductors with screened electron--electron interactions can exhibit
a strong rectification effect in the presence of spatially asymmetric 
impurity scatterers. 
The rectification current 
shows a power--law behavior, $I_r \sim V^z$, with an exponent
$z$ that depends on the electron interaction strengths $g_c$ and
$g_s$ only, and which can be determined from second and third
order perturbation theory. For small values of $g_c$ the exponent
becomes negative, leading to the upturn of $I_r$ about the voltage $V^*$
(Fig.~\ref{fig:upturn}).
We note that this effect was obtained in the framework of 
weak coupling theory. For voltages $V<V^*$, the effective
impurity scattering strength exceeds $E_F$, marking a crossover
into the strong backscattering regime\cite{Kane92,Feldman05}. 
All currents eventually vanish at $V=0$.

% --------------------------------------------------------------------------------------------------------

\acknowledgments
This work is supported in part by the NSF under grant
number DMR--0213818 and the Swiss National Fonds.

% --------------------------------------------------------------------------------------------------------

\appendix

% --------------------------------------------------------------------------------------------------------

\section{Estimate of higher perturbative orders}

In the preceding study we have shown that third order contributions can exceed
the second order ones for small values of $g_c,g_s$. 
To complete this, we have to give an additional proof that higher perturbative
orders $N \ge 4$ cannot exceed these values in the physical range of 
$g_c > 0$ and $g_s \ge 1/2$.

Since the calculation of the rectification current involves 
contributions that are lesser relevant than the leading 
power--laws for the backscattering current, we have also to show that
the neglected backscattering term in $\mathcal{S}_0$,
proportional to $\int \dx \cos(2 \phi_s)$,
cannot generate additional important corrections to the 
rectification current. 
Under renormalization, this term has to be completed
by allowing the general interaction 
$\sim \int \dx \cos(2 \hat{n}_s \phi_s)$ with a summation over 
integer $\hat{n}_s \neq 0$ (the notation $\hat{n}_s$ is used to distinguish
these indices from those $n_s$ appearing in $U(n_c,n_s)$).
Under the renormalization group, 
when integrating over high energies down
to the characteristic energy $V$, such terms are renormalized 
as\cite{Giamarchi04} $(V/E_F)^{(2\hat{n}_s)^2 g_s - 2}$.

We choose the following strategy: Let us assume to be in 
the perturbative region with voltages $V > V^*$, such that
$U,V \ll E_F$ and
$U (V/E_F)^{g_c n_c^2 + g_s n_s^2 - 1} \ll E_F$. 
A general higher order correction to the current has the form 
\begin{equation} \label{eq:INM}
	I^{(N,M)} = 
	V
	\prod_{\substack{\{n_c,n_s\}\\ N \ \text{factors}}}
	\bigl(
	U V^{g_c n_c^2 + g_s n_s^2 - 1}
	\bigr)
	\prod_{\substack{\{\hat{n}_s\}\\ M \ \text{factors}}}
	V^{(2\hat{n}_s)^2-2},
\end{equation}
where for simplicity we set $E_F = 1$ and neglect prefactors of
order 1. 
Charge and spin conservation here requires that 
$\sum n_c = 0$ and $\sum n_s + \sum (2 \hat{n}_s) = 0$.
Since $g_s \ge 1/2$, the product over the $V^{(2\hat{n}_s)^2-2}$
is always $\le 1$, and if we denote by $I^{(N)}$ the 
expression obtained from Eq.~\eqref{eq:INM} by suppressing
these $M$ factors, we always have $I^{(N,M)} \le I^{(N)}$.
Therefore, the corrections arising from $\int \dx \cos(2\phi_s)$ are always small.

Let us choose a subset of factors of $I^{(N)}$, and denote this quantity by $J$,
\begin{equation}
	J = V \prod_{\substack{\text{selection of }\{n_c,n_s\}}}
	\bigl(
	U V^{g_c n_c^2 + g_s n_s^2 - 1}
	\bigr).
\end{equation}
Since all factors are smaller than 1, we have
\begin{equation}
	I^{(N,M)} \le I^{(N)} < J.
\end{equation}
We will now show that, for $N \ge 4$, it is always possible to choose
a $J$ that is smaller or equal to the dominating second or third order 
contribution. The larger the $n_{c,s}$, the smaller become the factors 
in $I^{(N)}$ and $J$. Much in the analysis, therefore, depends on 
the maximal value of $|n_{c}|$, which we denote by $\bar{n}_{c}$.

This method of comparison does not hold for the comparison between
the second and third order expressions themselves. Since these
expressions are products of two or three factors only, we cannot choose a suitable
subset for $J$ but have to deal with the full expressions. 
The exponents are therefore largely determined by the constraints
of particle conservation plus having even sums of charge and
spin numbers. The result of comparison was discussed above. 
For $N \ge 4$, however, the larger freedom of choice of $J$
allows us to show that these higher order expressions are small compared
to the second and third order ones. 
The following estimates contain the proof of this statement.

For any choice of maximal $\bar{n}_c$  we
choose a suitable $J$ consisting of two or three factors
$U V^{n_c^2 g_c + n_s^2 g_s - 1}$ and the prefactor $V$ or,
if required, the prefactor $V^2$ with the additional $V$ arising
from the expansion of a $U$ (similarly to the second order 
calculation). We then show that this $J$
is smaller or equal to one of the expressions
\begin{align}
	I_A &= U^2 V^{2 g_c + 2 g_s},\\
	I_B &= U^2 V^{8 g_c},\\
	I_C &= U^3 V^{6g_c + 2g_s-2},\\
	I_D &= U^3 V^{24g_c - 2},
\end{align}
for any $g_c \ge 0$ and $g_s \ge 1/2$. 
This will prove that $I^{(N)} < J$ cannot exceed the dominant
second or third order expression, given by the maximum of
$I_{A,B,C,D}$.

Since $\bar{n}_c \ge 1$, the particle conservation imposes that
another or several numbers $n_{c}$ are different from zero.
These are denoted by $\tilde{n}_c$ and $\tilde{m}_c$ below.
If not further specified we only assume that $|\tilde{n}_c|,|\tilde{m}_c| \ge 1$.
If $\bar{n}_c$ is odd there is at least another odd $|\tilde{n}_c| \ge 1$ and, 
due to the requirement of even $n_c+n_s$, there exist at least two odd $|n_s| \ge 1$, 
denoted by $\tilde{n}_s$ and $\tilde{m}_s$.

We distinguish between the following five cases:

\begin{enumerate}[1)]
\item \emph{$\bar{n}_c \ge 2$ is even; all nonzero $|n_c| = \bar{n}_c$}:
If in the couples $(n_c,n_s)$ all $n_c \neq 0$ have $n_s = 0$, 
$I^{(N)}$ (and $I^{(N,M)}$) is symmetric under the change of sign
$V \to -V$, and the rectification current vanishes (note that the $n_s$ for which 
$n_c=0$ and the $\hat{n}_s$ can be nonzero though). 
This situation is similar to the second order case discussed 
above. A rectification current exists since the potential $U$ 
depends on the voltage through the density $k_F$. An expansion
of a $U$ to linear order in $V$ allows us to choose a $J$
of the form
\begin{equation} \label{eq:J1}
	J = V^2 (U V^{\bar{n}_c^2g_c - 1} )(U V^{\bar{n}_c^2g_c - 1} )
	\le U^2 V^{(2^2+2^2)g_c} = I_B.
\end{equation}
On the other hand, if not all $n_s$ are zero, there
must be an even $n_s = \tilde{n}_s$ with $|\tilde{n}_s| \ge 2$, and we can choose for $J$
\begin{multline} \label{eq:J2}
	J = V 
	(U V^{\bar{n}_c^2g_c + \tilde{n}_s^2g_s - 1})
	(U V^{\bar{n}_c^2g_c - 1})
	\\
	\le
	U^2 V^{(2^2+2^2)g_c + 2^2 g_s - 1}
	\le
	U^2 V^{8g_c + 2 g_s}
	\le I_{A,B},
\end{multline}
where we have used $g_s \ge 1/2$.
\item \emph{$\bar{n}_c \ge 2$ is even; all nonzero $|n_c|$ are even but are not all equal}:
This condition excludes $\bar{n}_c = 2$ as it coincides with the previous case.
For $\bar{n}_c \ge 4$ there exist two
other nonzero and even $|\tilde{n}_c|,|\tilde{m}_c| \ge 2$ 
or another $|\tilde{n}_c| \ge 4$ and an arbitrary $|\tilde{m}_c| \ge 0$.
We then choose
\begin{equation}
	J 
	= U^3 V^{(\bar{n}^2+\tilde{n}_c^2+\tilde{m}_c^2)g_c-2}
	\le U^3 V^{(4^2+8)g_c - 2} = I_D.
\end{equation}
\item \emph{$\bar{n}_c \ge 2$ is even; there is an odd $|n_c|$}:
Since $\sum n_c = 0$ there exist at least two odd 
$|\tilde{n}_c|,|\tilde{m}_c| \ge 1$, and since $n_c+n_s$ must be even,
there must be two odd $|\tilde{n}_s|,|\tilde{m}_s| \ge 1$.
This leads to the bound
\begin{multline}
	J =
	U^3 V^{(\bar{n}_c^2+\tilde{n}_c^2+\tilde{m}_c^2)g_c + (\tilde{n}_s^2+\tilde{m}_s^2)g_s-2}
	\\
	\le
	U^3
	V^{(2^2+1+1)g_c + (1+1)g_s - 2}
	= I_C.
\end{multline}
\item \emph{$\bar{n}_c = 1$}:
In this case all $n_c$ are $|n_c|=1$ or zero. If all nonzero $n_c$ have
$n_s = \pm 1$, $I^{(N,M)}$ is symmetric under $V\to -V$, and we have to 
expand a $U$ to linear order in $V$. If we choose a $|\tilde{n}_c| = 1$ 
and two $|\tilde{n}_s|,|\tilde{m}_s| = 1$, we can set
\begin{multline}
	J = V U^2 V^{(\bar{n}_c^2 + \tilde{n}_c^2)g_c + (\tilde{n}_s^2+\tilde{m}_s^2)g_s -1}
	\\
	\le U^2 V^{(1+1)g_c + (1+1)g_s} = I_A.
\end{multline}
On the other hand, if there is a couple $(\tilde{n}_c,\tilde{n}_s)$ with 
$\tilde{n}_c \neq 0$ and $|\tilde{n}_s| \ge 3$, 
we have the estimate
\begin{multline}
	J = U^2 V^{(\bar{n}_c^2+\tilde{n}_c^2)g_c + \tilde{n}_s^2 g_s - 1}
	\le U^2 V^{(1+1)g_c + 3^2 g_s - 1}
	\\
	\le U^2 V^{2g_c+7g_s}
	\le I_A.
\end{multline}
\item \emph{$\bar{n}_c\ge 3$ is odd}:
Since $\bar{n}_c$ is odd there must be at least another odd $|\tilde{n}_c| \ge 1$,
and, to fulfill that $n_c + n_s$ is even, 
there must be two odd $|\tilde{n}_s|, |\tilde{m}_s| \ge 1$, such
that
\begin{multline}
	J 
	= 
	U^2 V^{(\bar{n}_c^2+\tilde{n}_c^2)g_c+(\tilde{n}_s^2+\tilde{m}_s^2)g_s - 1}
	\\
	\le
	U^2 V^{(3^2+1)g_c + (1+1)g_s-1}
	\le
	U^2 V^{10g_c} 
	\le I_B.
\end{multline}
\end{enumerate}
We conclude that higher order terms cannot exceed
the second and third order expressions for $V \gtrsim V^*$.

% ----------------------------------------------------------------------------

% ----------------------------------------------------------------------------
%

\pagebreak

\begin{table}
\begin{center}
\begin{tabular}{l}
2\textsuperscript{nd} order:
\\
\begin{tabular}{|c|c|r r r r|}
\hline
phase & exponent $z$ & $n_c$ & $n_s$ & $m_c$ & $m_s$  \\
\hline
(A) & $2g_c+2g_s$	& $1$ & $1$ & $-1$ & $-1$ \\
(B) & $8g_c$ 		& $2$ & $0$ & $-2$ & $0$  \\
\hline
\end{tabular}
\\
3\textsuperscript{rd} order:
\\
\begin{tabular}{|c|c|r r r r r r|}
\hline
phase & exponent $z$ &  $n_c$ & $n_s$ & $m_c$ & $m_s$ & $l_c$ & $l_s$  \\
\hline
(C) & $6g_c+2g_s-2$	 & $1$ & $1$ & $1$ & $-1$ & $-2$ & $0$\\
(D) & $24g_c-2$		 & $2$ & $0$ & $2$ & $0$ & $-4$ & $0$\\
\hline
\end{tabular}
\end{tabular}
\end{center}
\caption{Most relevant exponents $z$ for the rectification current
(see Eqs.~\eqref{eq:z2} and \eqref{eq:z3}).
Permutations and sign changes of $n,m,l$ leading to the same exponents are not 
shown. 
\label{tbl:z}}
\end{table}

% ----------------------------------------------------------------------------

\end{document}